\documentclass[fleqn,usenatbib,utf8]{mnras}
\usepackage[T1]{fontenc}

\DeclareRobustCommand{\VAN}[3]{#2}
\let\VANthebibliography\thebibliography
\def\thebibliography{\DeclareRobustCommand{\VAN}[3]{##3}\VANthebibliography}

\usepackage{graphicx}	
\usepackage{amsmath}	
\usepackage{amssymb}	
\usepackage{datetime2}
\usepackage{graphicx}
\usepackage{booktabs}
\usepackage{tikz}
\usepackage{enumitem}
\usepackage{siunitx}

\usepackage[]{hyperref}

\usepackage[capitalise]{cleveref}

\usepackage{newtxtext,newtxmath}

\usepackage[utf8]{inputenc}
\newcommand{\ntild}{\char`\~}
\newcommand\so{\mathring{s}}
\newcommand\So{\mathring{S}}
\newcommand\tto{\mathring{t}}

\newcommand\Vop[2]{\mathcal{V}\left\{#1,#2\right\}}
\newcommand\lam{(\lambda)}
\newcommand\lamp{(\lambda_i')}
\newcommand\alam{(a\lambda)}

\title[Stellar Karaoke]{Stellar Karaoke: Deep Blind Separation of Terrestrial Atmospheric Effects out of Stellar Spectra by Velocity Whitening}

\author[Nima Sedaghat et al.]{Nima Sedaghat,$^{1,2}$\thanks{E-mail: nimaseda@uw.edu (NS)}
Brianna M. Smart,$^{1}$\thanks{These authors contributed equally to this work}
J. Bryce Kalmbach,$^{1}$\footnotemark[2]
Erin L. Howard$^{1}$ and
Hamidreza Amindavar$^{3}$
\\
$^{1}$DiRAC Institute and the Department of Astronomy, University of Washington, Seattle, WA, U.S.A\\
$^{2}$The Raw Data Speaks Initiative\\
$^{3}$Department of Electrical Engineering, Amirkabir University of Technology, Tehran, Iran
}

\date{Accepted 2023 August 22. Received 2023 August 22; in original form 2023 January 17}

\pubyear{2023}

\begin{document}
\label{firstpage}
\pagerange{\pageref{firstpage}--\pageref{lastpage}}
\maketitle

\begin{abstract}
We report a study exploring how the use of deep neural networks with astronomical Big Data may help us find and uncover new insights into underlying phenomena:
through our experiments towards unsupervised knowledge extraction from astronomical Big Data we serendipitously found that deep convolutional autoencoders tend to reject telluric lines in stellar spectra. With further experiments we found that only when the spectra are in the barycentric frame does the network automatically identify the statistical independence between two components, stellar vs telluric, and rejects the latter.
We exploit this finding and turn it into a proof-of-concept method for removal of the telluric lines from stellar spectra in a fully unsupervised fashion: we increase the inter-observation entropy of telluric absorption lines by imposing a random, virtual radial velocity to the observed spectrum. This technique results in a non-standard form of ``whitening'' in the atmospheric components of the spectrum, decorrelating them across multiple observations .
We process more than 250,000 spectra from the High Accuracy Radial velocity Planetary Search (HARPS)  and with qualitative and quantitative evaluations against a database of known telluric lines, show that most of the telluric lines are successfully rejected.
Our approach, `Stellar Karaoke', has zero need for prior knowledge about parameters such as observation time, location, or the distribution of atmospheric molecules and processes each spectrum in milliseconds.
We also train and test on Sloan Digital Sky Survey (SDSS) and see a significant performance drop due to the low resolution.
We discuss directions for developing tools on top of the introduced method in the future.
\end{abstract}

\begin{keywords}
techniques: spectroscopic -- atmospheric effects -- methods: data analysis -- methods: statistical
\end{keywords}

\maketitle

\begin{figure}
  \begin{tikzpicture}
    \node[anchor=south west,inner sep=0] at (0,0) {\includegraphics[width=\linewidth]{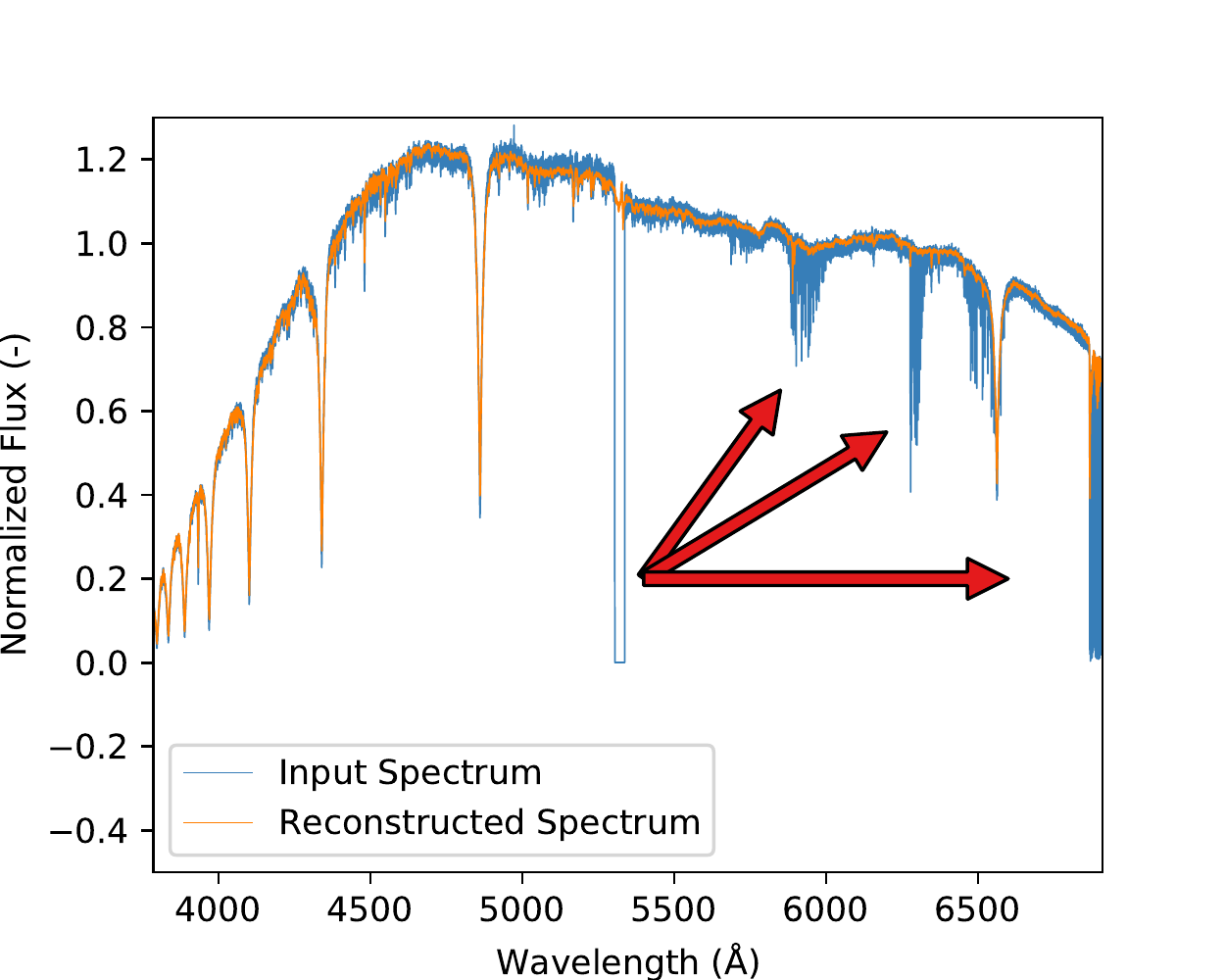}};
    
  \end{tikzpicture}
  \caption{We exploit the statistical properties of stellar spectra in large datasets, pass them through a convolutional autoencoder, and get telluric lines rejected with minimal effort. The three red arrows in the figure highlight spectral lines that the autoencoder identified as telluric lines and did not include in the reconstructed spectrum.}
  \label{fig:teaser}
\end{figure}

\section{Introduction}
Throughout this paper, we introduce and elaborate a method based on a novel signal processing-based technique combined with a fully unsupervised\footnote{A training scheme that does not require ground truth data -- see \citet{james2023unsupervised} for an overview.} deep neural network to remove the non-linear, time-variant effect of telluric lines out of observed spectra without any need for manual modelling and tuning -- \cref{fig:teaser}.
Concretely, we show how looking at an enormous dataset of spectra reveals new insights into their features, allowing for novel solutions to existing, difficult problems.

Through our experiments towards unsupervised knowledge extraction from astronomical Big Data, we trained deep neural networks for non-linear dimensionality reduction of a large number of archival stellar spectra and serendipitously found that the networks tend to drop (\cref{fig:teaser}) specific regions in the reconstructed version of the spectra: the telluric lines. With further experiments we found that only when the spectra are in the barycentric frame, the network shows this behaviour. A ``barycentric frame'' is a reference frame centered on the center of mass of a celestial system, providing a stable reference for high-precision astronomical measurements \citep{soffel2003iau, bohlin2014techniques}. Contrastingly, a "topocentric frame" is centered on an observer's location on Earth. The latter, which is the natural coordinate system in the as-observed representation of spectra, does not account for the movement of Earth around the Solar System's barycenter, and thus may introduce variability or inaccuracies into measurements of stellar spectra, often calling for transformation to the barycentric frame.

Our further experiments showed that, by looking at a large number of such transformations, the network automatically identifies the statistical independence between two components, stellar vs telluric, and decides to reject the latter. 
We exploit this finding and turn it into a proof-of-concept method for removal of the telluric lines from a whole set of stellar spectra in a fully unsupervised fashion: we increase the inter-observation entropy of telluric absorption lines by imposing a random, virtual radial velocity to the observed spectrum, in the original, topocentric frame. This technique results in a non-standard form of ``whitening'' in the atmospheric components of the spectrum, decorrelating them across multiple observations -- similar to what transformation to the barycentric frame would do, but with zero need for observation-related parameters.

\subsection{Atmospheric Effects on Stellar Spectra}
In ground-based observations of stellar spectra, it is not unusual for the source's spectrum to reach our detectors in an altered state. The spectrum may be affected by a number of events in space (redshift/blueshift, emission, absorption, etc.). When finally reaching the Earth, the photons in the spectrum will undergo additional transformations as they travel through the Earth's atmosphere and are collected by telescopes. Telescope effects are often well characterised and constrained (CCD noise, fringing, mirror defects, etc.). However, other effects are usually more difficult to model and correct.

Particularly, passing through the terrestrial molecules present in the Earth's atmosphere results in absorption lines in the observed spectrum. These absorption lines, known as telluric lines, are the results of interactions between photons and molecules due to a number of electron, rotational, and vibrational transitions \citep{rybicki1986radiative}. These interactions are a persistent source of contamination and loss of information in the observed spectra -- e.g. see \cite{bolton2012spectral}, \cite{artigau2014telluric}.
The majority of the molecules responsible for absorption are $\textrm{O}_{2}$, $\textrm{H}_{2}\textrm{O}$, $\textrm{CO}_{2}$, $\textrm{CO}$, $\textrm{CH}_{4}$, $\textrm{COH}$, $\textrm{O}_{3}$, $\textrm{N}_{2}\textrm{O},$ and the Chappuis ozone absorption bands \citep{catanzaro1997high, griffin2005detection, moehler2014flux, smette2015molecfit}. Many of these absorption features lie in the near-infrared and ultraviolet part of the electromagnetic spectrum, with weaker absorption features from ozone, oxygen, and water present in optical wavelengths. Additional emission features are also added to the spectrum as photons are emitted from the molecules. The emission features are relatively straightforward to handle by observing source-free portions of the sky. Observed emission lines in "empty" regions provide an easily accessible template for emission removal. The absorption features are more complicated. Different telluric lines can scale linearly or non-linearly dependant on the airmass, and are affected by atmospheric conditions at the time of observation \citep{bailey2007infrared}.

Most of the traditional methods used for the removal of telluric lines consider each spectrum as a single independent entity, discarding the set of observed spectra as a whole -- e.g. see \cite{hrudkov2005reliable}. One group of such methods use standard stars to assist with atmospheric line removal. They do this by observing standard stars with relatively featureless spectra and using them as atmospheric templates. This is often done by observing A0V or G-type stars, though this is predominantly used at the near-infrared and IR wavelengths \citep{vacca2003method,artigau2014telluric}. When using G-type stars, high resolution Fourier transform of spectra of the sun can be used as templates to correct for the intrinsic lines in the star \citep{livingston1991atlas,hase2010acefits}.

The reference star should be ideally located near the target star and observed close in time to measure the atmospheric conditions as accurately as possible. The target source's spectrum is then divided by the telluric template. However, this method is limited if there are no such standard stars available during an observing run or if observing conditions result in a poor spectrum. Atmospheric absorption can also vary rapidly, on the order of minutes \citep{moehler2014flux}, requiring significant time and resources dedicated to observing the template stars. The division is an iterative process requiring wavelength and intensity scaling adjustments to match the atmospheric effects.

Another common method is to model the Earth's atmosphere and create a synthetic spectrum which precisely models the atmospheric absorption. These models solve the radiative transfer equation of the Earth's atmosphere using numerical models -- e.g. \cite{allart2022automatic}. These methods are dependent on atmospheric conditions measurements from the night's observations. This is commonly done with programs such as \texttt{molecfit} \citep{smette2015molecfit,kausch2015molecfit} and \texttt{telfit} \citep{ gullikson2014telfit}. The radiative transfer code retrieves the temperature, pressure, and humidity from the time of observation and uses a database of molecular parameters to create a fit for the telluric absorption. While this technique is relatively successful, it may perform poorly if there are a large number of intrinsic features, little or no continuum, low signal-to-noise ratio (S/N), or large airmass observations with high water vapour content (see \citet{alonsofloriano2019he,shulyak2019magnetic}). Moreover, current implementations of this technique, e.g. \texttt{molecfit}, suffer from rather slow performance to the extent that fitting a model to a single spectrum may take up to several minutes on today's computers. \cite{mcloat2021exoplanets,nicholls2017crires} report on other types of model-fitting problems that \texttt{molecfit} encounters  in their particular experiments. See \cite{ulmer2019telluric} for a comparison of the above methods.

\subsection{Dimensionality Reduction}
Some recent lines of work in astronomy have used implicit modelling of the spectra by looking at a set of examples, potentially eliminating the need for manual, explicit modelling. Such methods, loosely dubbed data-driven approaches, have gained a good momentum in the past couple of decades. Particularly in astronomy, a data-driven look at spectra has been mainly utilised in the broad context of dimensionality reduction.

Arguably the most popular dimensionality reduction method used in astronomy has been the Principal Component Analysis (PCA -- see \cite{jolliffe2016principal} for a review), which has been applied on astronomical spectra for many years now. \cite{connolly1995PCA} used it to classify galaxy spectra with only the first two principal components while \cite{bailer1998automated} showed that PCA can compress stellar spectra by a factor of over 30 and classify anomalous and non-stellar spectra in the dataset. Furthermore, \cite{bailer1998automated} found that the compression removed noise as well as bogus features in the spectra such as dust appearing on the plate during plate scanning. Some works such as \cite{artigau2014telluric}, even attempt to remove telluric lines of a few specific objects in the High Accuracy Radial velocity Planetary Search project (HARPS; \citealp{2003Msngr.114...20M}) dataset using PCA decomposition and demonstrate improved radial velocity measurement accuracy. However, their method does not operate solely on the stellar spectra and still requires observations of telluric standard stars. Our approach solely requires a set of input stellar spectra and zero additional information.

Moreover, the very important, but often overlooked point about applicability of PCA for such use-cases, is its linear nature, making it unable to account for modelling non-linear phenomena by definition. E.g. in the case of telluric lines, a method simply based on PCA, may not be able to tell the difference between a telluric and stellar line, in \textit{busy} regions of the spectrum. This inability has been well observed as early as in \cite{bailer1998automated}. In this work, we show how transformation to a more sophisticated feature space is necessary for such a task.

\subsection{Autoencoders}
\label{sec:intro-auto-encoders}
There is a closely related family of neural networks known as encoder-decoder networks (\cref{fig:architecture1}).
The \textit{contraction} part compresses the input to an often low-dimensional representation, also known as ``latent representation'' or simply a ``code''. The \textit{expansion} part is then used to \textit{decode} this low-dimensional representation up to the desired output. The transformation used for coding (and decoding) is learnt during training, according to the task at hand. A special case of such networks, an \textit{autoencoder}, is an unsupervised network trained to reconstruct the input without the need for labelled input data. As described in \cite{hintonAutoencoderDimensionality} these networks can be used to reduce the dimensionality of input data in a non-linear generalisation of PCA.  \cite{wang2016auto} shows comprehensive experiments on dimensionality reduction using autoencoders, and studies the effects of different latent dimensions using  synthetic and real images. 

In \cite{yang2015autoencoder} the authors use a classical (non-convolutional) autoencoder to transform 3000-dimensional spectra from the Sloan Digital Sky Survey (SDSS) Data Release 7 \citep{sdssDR7} to lower-dimensional features, which are later used for estimation of atmospheric parameters.
\cite{wang2016new} use semi-classical (fully-connected decoder) autoencoders for feature learning on astronomical spectra. They compare spectral classification based on their learnt features with PCA and locally linear embedding (LLE), an alternate non-linear dimensionality reduction method. They find that their autoencoder approach performs the best when classifying spectra among a dataset of F, G and K-type stars.

One problem with the mere use of autoencoders, especially for applications in (astro-)physics, is the entangled mapping of concepts into the reduced signal -- the latent space. The Variational AutoEncoder (VAE; \citealp{kingma_auto-encoding_2014}) has shown to mitigate this effect to some extent. \cite{portillo2020sdssVAE} used a VAE to reconstruct SDSS galaxy spectra re-sampled to an input spectrum with 1000 components. They found that when limiting the dimensionality of the latent space to 2, 4, 6 or 10 components ($\geq 99\%$ compression) the VAEs and traditional autoencoders both reconstructed SDSS galaxy spectra with a lower reconstruction error than PCA and non-negative matrix factorisation (NMF) applied with the same number of components. The authors then demonstrated that different galaxy classes occupied different areas of the latent space enabling classification.

Classical neural networks (including autoencoders) are merely composed of fully-connected layers \citep{hinton_reducing_2006}. But on the one hand, with the advent of deep neural networks we have learnt that having deeper networks is essential for learning more useful features \citep{szegedy2015going}, and on the other hand, fully-connected networks suffer from a lack of scalability to deeper networks and high-dimensional data \citep{bengio2009learning}. As such, a deep enough autoencoder merely based on fully-connected layers is not tractable for modern spectroscopic signals: e.g. with $\sim$~300,000 pixels in case of HARPS, a fully-connected autoencoder composed of merely 6 layers has more than 6.5 billion trainable parameters, which does not fit into the memory of a normal GPU of today, even with a single spectrum as input. See \cref{app:fullyconnected} for an example of such an architecture. Therefore, stellar spectra have usually been down-sampled to \textit{tractable resolutions}: 1000, 2000 and 3000 pixels in \citealp{portillo2020sdssVAE}, \citealp{wang_auto-encoder_2016} and \citealp{yang2015autoencoder}, respectively. This results in irrecoverable loss of high-resolution information content in originally high-resolution spectra, as we will show in one of our experiments too. We address this issue by additionally making use of convolutional and up-convolutional layers for transformation of data to the desired feature space; an idea borrowed from computer vision, which also allows the network to obtain a degree of translation invariance \citep{fukushima1980neocognitron}. This way we manage to train an autoencoder as deep as 32 layers with less than 6 million trainable parameters, which is now tractable with today's hardware and gives us new insights in to stellar spectra.
Further study of technicalities of convolutional layers, their advantages and choice of hyperparameters is beyond the objectives of this paper; but the interested reader may refer to the computer vision literature of the two first decades of the current millennium (including the above mentioned works) to obtain an elaborate picture of this transition from classical machine learning to deep learning.

Note that although \textit{convolutional} neural networks have been vastly applied on astronomical data in the past years, they have mostly focused on extracting information from 2D images: from photometric redshift regression \citep{hoyle2016photozCNN, pasquet2019photozCNN} and galaxy morphology classification \citep{deileman2015galaxyCNN, huertasCompany2015galaxyCNN, ghosh2020galaxyCNN} to image-generating encoder-decoders for transient detection \citep{sedaghat2018effective}.
However, 1D convolutional networks, necessary for work on spectra, have been rather rarely used in astronomy. \cite{sharma2020spectraCNN} implement such a network to classify stellar spectra from SDSS and compare their method to a fully connected (feed-forward) implementation.

\subsection{Blind Source Separation}
Telluric lines inherently originate from a source different to that of stellar lines. This makes their elimination a ``source separation'' problem; a topic mainly studied in digital signal processing. The general definition of the problem assumes multiple sensors and multiple signals -- a Multiple Input Multiple Output (MIMO) system. In the problem at hand, however, we only have a single output, making the task slightly more difficult. Moreover, most of the classic formulations of source separation consider linear mixing of signals from various sources, whereas in our case we are far from such a simple configuration. Due to the random nature of the radial velocity in each observed spectrum, and the fact that we do not make any assumptions about the nature of the effects \textit{after} the mixing of the two signals has happened, it can be seen as a ``blind'' source separation task (BSS).

Recent BSS methods have typically centreed around Independent Component Analysis \citep{comon1992independent} and its extensions -- e.g. see \citet{choi2005blind,pal2013blind} for a review. Denoising Source Separation (DSS), introduced by \citet{sarela2005denoising}, is another approach that casts the task of source separation as a denoising procedure, making it similar to the basic concept utilised in our method. But DSS aims for statistical estimation of the \textit{mixing matrix} (the mixing mechanism), whereas the objective in our method is to directly recover the desired signal by whitening the unwanted component, making it more noise-like.

Many recent methods of source separation, particularly for audio/music signals, have utilised encoder-decoder neural networks as the canonical component of their methods \citep{kameoka2018semi,seki2019underdetermined,grais2017single,liu2018denoising}. But they mainly rely on explicit supervision calling for availability of synthetic ground truth. \cite{neri2021unsupervised} train a variational encoder-decoder network, where they rely on disentanglement properties of the VAE to produce separated sources in the output. The mentioned audio-processing methods all train 2D  neural networks on spectrogram \textit{images}. \cite{wisdom2020unsupervised} introduce a semi-supervised audio source separation method, working directly on the time-domain signals. Although they implement both supervised and self-supervised\footnote{A training scheme which relies on ground-truth data that comes ``for free'': the labels are automatically obtained from the unlabelled input data.} networks, the authors show that the latter should be regularised by the former to work properly. The unsupervised part of the last two papers relies on the strong assumption that the input is a linear mixture of its constituent sources. Our network is a simple autoencoder with no linearity assumptions and proves to work well in the complex non-linear conditions formulated in the upcoming sections. In this context, our Karaoke network sees the stellar spectrum as the music, and the adversarial effects as the vocal to be removed from it.
   
\subsection{Composite Spectra}
Another line of research towards disentanglement of composite spectra has existed as early as 1994, when \citet{simon1994disentangling} try to decompose the constituent components of spectra of binaries. More recently, tools like \texttt{wobble} \citep{bedell2019wobble} have similarly modelled the stellar and telluric parts as two separate components and have tried to fit these parametric models to the observed spectra. Such methods, however, rely on availability of parameters of observation time, the main of which being BERV: the Barycentric Earth Radial Velocity. More importantly, they require several observations of the same star in a single data collection, to obtain a diverse enough set of BERV values for a single object, which is not usually available for too many objects. These models are also object-specific, in the sense that each model is prepared (i.e. optimised) for a single object and is dedicated to processing of spectra from that object only.

Our method, conversely, does not require any external information other than the spectra themselves: a raw set of scalar vectors. Although the underlying physical mechanism that justifies the usage of different BERV values is close to that of our whitening technique, our method relies on the diversity across several spectra of different objects, while at the same time captures the common patterns in them, and therefore is not limited to specific objects -- neither during the training phase, nor when being applied on spectra. We also provide performance comparison to \texttt{wobble} in our experiments section.

Perhaps the closest to our work, both in terms of the used network architecture, as well as the data-driven unsupervised concept is \cite{sedaghat2021machines}, where a convolutional VAE is used to extract knowledge from a large number of HARPS spectra, in a fully unsupervised manner. We borrow and use the exact same neural architecture for our work. 

\paragraph*{\textbf{Our Contributions}}
\begin{itemize}[label=\textbullet]
    \item We process a huge number of very high-dimensional spectra as a whole, letting statistical properties emerge in them, allowing to be treated as random processes.
    \item We model the telluric components in stellar spectra as independent stars and impose a virtual radial velocity on them to achieve statistical whitening/decorrelation.
    \item We use a convolutional autoencoder, that automatically acts as a source separation tool, rejecting telluric lines in a fully unsupervised fashion: zero explicit modelling and zero annotation efforts.
    \item Our method is object-independent and processes new observations in a fraction of a second.
\end{itemize}
\section{Problem Formulation}

\begin{figure*}
  \begin{center}
    \includegraphics[width=\linewidth]{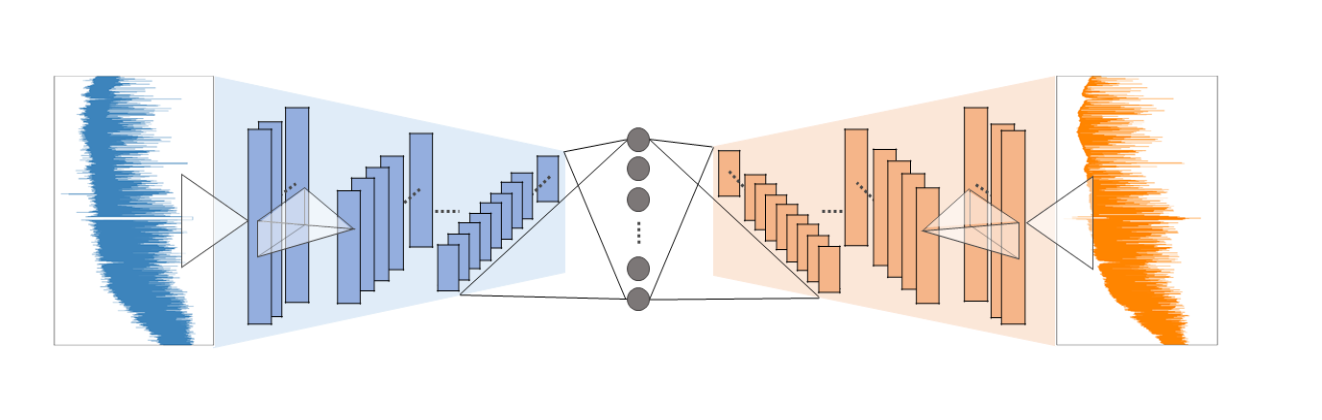}
  \end{center}
  \caption{A typical variational autoencoder trained to reconstruct stellar spectra, can decompose physically meaningful components out of the input, when the compression is high enough and the number of convolutional kernels is kept low. 
  The variational implementation of the bottleneck is omitted for the sake of simplicity. 
  The main contraction and expansion parts on the left and right are composed of (up)convolutional layers, while fully connected layers are necessary at the bottleneck.  
  Blue is the input spectrum and orange is the reconstructed version. This image has originally been published in \citet{sedaghat2021machines}.}
  \label{fig:architecture1}
\end{figure*}

\label{sec:formulation}
We seek to clean adversarial atmospheric effects out of an arbitrary observed signal, $x$. The observed signal in our case is a stellar spectrum, and so is a function of wavelength, $\lambda$, letting us denote it as $x(\lambda)$.

We use the below formulation to model the various phenomena affecting a stellar spectrum, before it is captured by our sensors:

\begin{equation}
\begin{split}
    x(\lambda) = \Big( s(\lambda)   t(\lambda) \Big)*h(\lambda) + n(\lambda)
\end{split}
\end{equation}

\noindent where $s$ is the stellar spectrum, incorporating line-of-sight effects from relative stellar velocity and the interstellar medium. $t$ is an imaginary signal representing the telluric lines affecting the spectrum, also referred to as the atmospheric transmission spectrum \citep{rudolf2016modelling}.

We use $h$ to denote what we call the observation transfer function, which models the changes the signal goes through during the sensing process, and includes, but is not limited to, the instrumental profile \citep{rudolf2016modelling}.
$n$ models the additive noise which is not modelled in the transfer function, $h$.

We work with a large ensemble of $N$ observations, mostly coming from different sources. We use the subscript, $i$, to differentiate between various observations:

\begin{equation}
\begin{split}
    x_i(\lambda) = \Big( s_i(\lambda)   t_i(\lambda) \Big)*h(\lambda) + n_i(\lambda)
    \\
    i \in \{1,2,\dots,N\}
\end{split}
\end{equation}

\noindent Note that the spectrum, $s$, depends on $i$, as we work with various objects at the same time. The atmospheric conditions are also time-variant, an important point that is reflected in dependence of $t$ on $i$. The same is true for noise, $n$. Without loss of generality, and for the sake of simplicity, we assume that $h$ is constant across the whole set of observations.

We seek to eliminate the effect of telluric lines from the observed signal, which in this model is equal to extracting $s_i(\lambda)*h(\lambda)$ for every observation -- note that removal of the \textit{observation effect}, $h$, is not part of the objective here.

We also model the effect of the radial velocity, $v$, of the target object on the observed spectrum, $s$ as:

\begin{equation}
\label{eq:radvel}
\begin{split}
    s_i(\lambda) &= \Vop{\so_i(\lambda)}{v_i} \\
    &=\so_i\Big(\lambda\big(1-\frac{v_i}{c}\big)\Big)
\end{split}
\end{equation}

\noindent where $\so_i$ would be the observed spectrum, if the radial velocity was zero -- i.e. no Doppler shift. We call $\so_i$ the \emph{static spectrum} hereafter. The $\Vop{\cdot}{\cdot}$ operator represents the nonlinear
\footnote{The operator is linear with respect to its first argument, but nonlinear with respect to the second one: $\Vop{s}{v_1+v_2} \neq \Vop{s}{v_1}+\Vop{s}{v_2}$}
effect of the radial velocity, $v_i$, on the spectrum, as expanded in the second line of the above equation, and $c$ is the speed of light. The physical units of $v_i$ and $c$ can be arbitrarily chosen, as long as they are kept the same.

$t_i$ on the other hand, and by definition, do not have any dependence on $v_i$, and therefore can be modelled as:
\begin{equation}
\begin{split}
    t_i(\lambda) = \Vop{\tto_i\lam}{0}
\end{split}
\end{equation}

\subsection{Continuous vs Discrete}
\label{sec:discrete}
All the signals discussed above are of continuous nature up until the point they are sensed by the detector. Sensing by the detector is a process that involves sampling as one of its steps, converting a spectrum into a series of real-valued samples. 
Therefore, the data we work with in our experiments is the discrete representation of $x_i\lam$, namely $X_i^l$ -- we use $l$ as the discrete-valued index for the sampled pixels.

However, for the reasons below, keeping the formulation in the continuous representation is safe -- and clearer. First, the signals modelled so far are all the constituent elements of the pre-sampling signal, $x_i\lam$, and so the continuous models hold. Secondly, the only part of our method which explicitly modifies the signal along the wavelength axis, the $\mathcal{V}$ operator (\cref{sec:whitening}), does its job by regridding the interpolated version of  the signal; practically converting the discrete signal back to its continuous version, applying the transformation and sampling it back again to the discrete space. Therefore mathematical definition of the operator in the continuous space is valid.

\subsection{Signals as Random Processes}
\label{sec:random_process}
The process of sensing a signal, $x_i$, from an arbitrarily chosen object in the sky, which is affected by many non-deterministic phenomena along the way, can be seen as one realisation of a random process. In other words, the set of $x_i$, or their discrete representation, $X_i$, for various $i$, represent an ensemble of realisations of a random process, $\{X\}$\footnote{The term \textit{random} in this context is not in contradiction with the \textit{structure} in stellar spectra. The structure is encoded in the basic parameters of the constituent set of random variables, such as the expected value and covariance matrix.}.
Note that in this particular application, the index set of the random process is sampled from wavelengths, $\lambda$ -- a bit counter-intuitive, as it is usually of a temporal nature in typical applications. 

Therefore each $X^l$ (as defined in \cref{sec:discrete}) represents a random variable, with outcomes $X_{\pmb{i}}^l$. Similarly, we can model $\{S\}$, $\{\So\}$ and $\{T\}$ as discrete-index random processes of the same type for $s_i$, $\so_i$ and $t_i$, respectively. $\{S\}$ is then a generator of various \emph{clean} spectra, $S_i$, while $\{\So\}$ generates the static $\So_i$ spectra.
As we will see in the next section, this non-deterministic view on signals allows us to explain the statistical operations and properties of the components more clearly.

\begin{figure}
        
    \includegraphics[width=\linewidth]{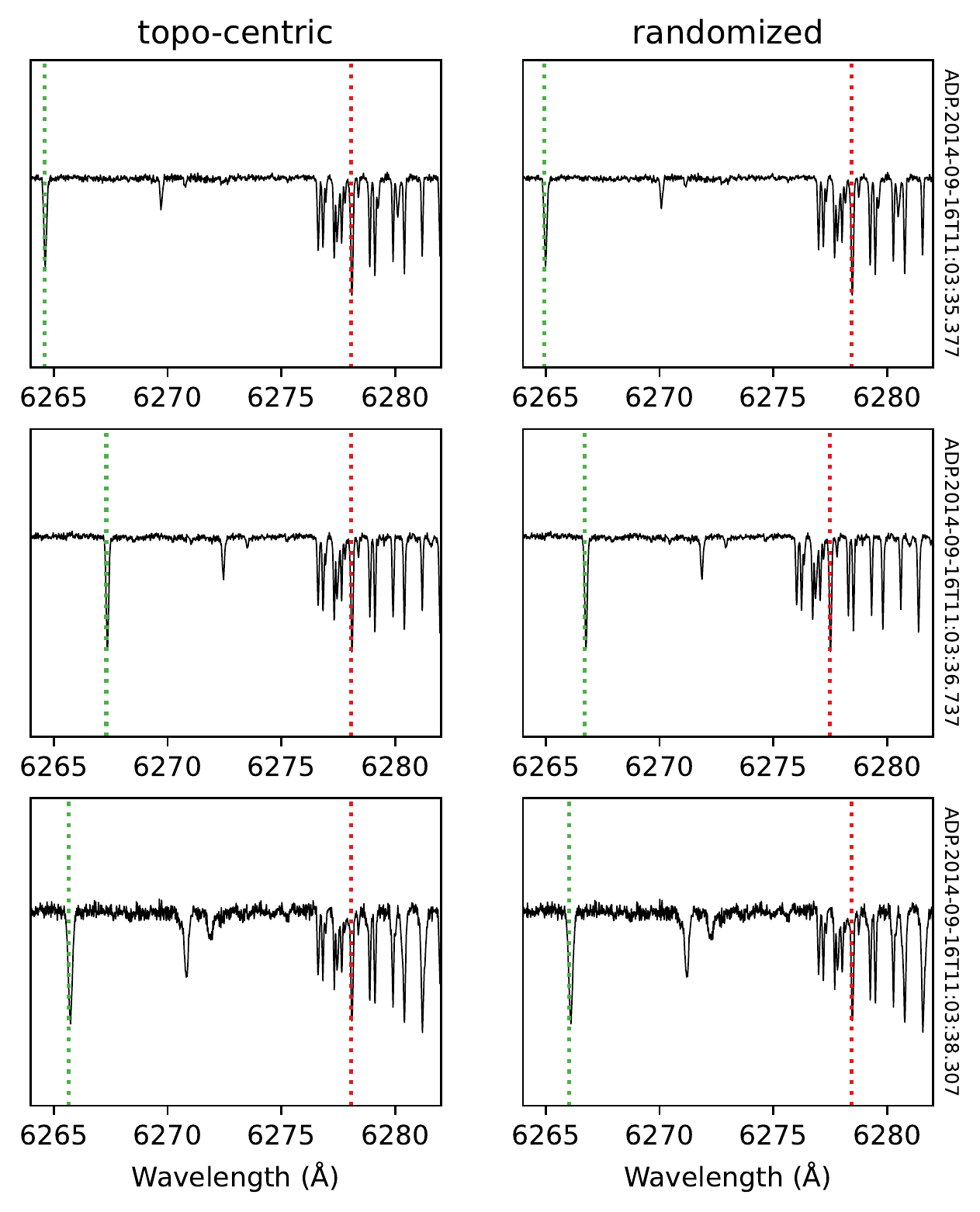}
    \caption{On the left a set of exemplar spectra are depicted. Stellar lines are unaligned due to different radial velocities. But telluric lines are aligned, even though they may have different shapes due to their inherent time dependence. On the right the same spectra after velocity randomisation are depicted. Telluric lines are now unaligned too, but with a pattern different to that of stellar lines. The red and green dotted vertical lines indicate the location of exemplar telluric and stellar lines, respectively. More telluric lines can be seen to the right of the marked one.}
    \label{fig:independence}
\end{figure}

\section{Method}
The method is composed of two key steps: 
\begin{itemize}
\item[]a) Increasing the entropy of telluric components across observations and,
\item[]b) transforming the spectra into a space where the stellar and telluric components are separable, then removing the non-dominant one.
\end{itemize}

\paragraph*{Median Normalisation}
As a pre-processing step, we normalise each spectrum in the dataset based on its \textit{median}, to mitigate the effect of different distances in sources, which otherwise results in extreme inter-sample flux range variations.

\begin{figure}
    \includegraphics[width=\linewidth]{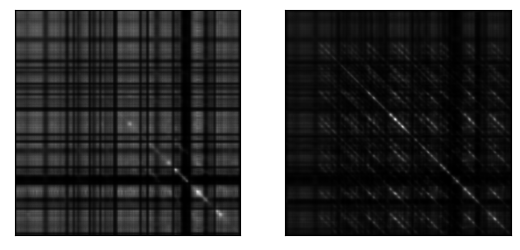}
    \caption{Visual illustration of the covariance matrix of the ensemble of signals, before and after whitening. On the left, the covariance matrix of the [6275, 6285] {\AA} region for some subset of size 90 of the observations is visualised. On the right, the same is done \textit{after} velocity whitening, where $v_i$ were sampled from the uniform distribution: $V \sim U(-30\text{ km/s, }30\text{ km/s})$. The covariance matrix is closer to the identity matrix now, confirming achievement of some degree of whitening/decorrelation.}
    \label{fig:whitening}
\end{figure}

\subsection{Velocity Whitening}
\label{sec:whitening}

In the default conditions, the spectroscopic observations are in the so-called topocentric reference frame where the telluric lines, if they exist, take on the same wavelengths -- or pixels locations. What it means from a statistical point of view though is the following:

Let us assume, for simplicity, that all the observed objects have the same static spectra, $\so_i\lam$. The resulting spectra, $s_i\lam$ would then only be different based on their radial velocities, $v_i$, which, by definition, are modelled as scaling transformations along the wavelength axis (\cref{eq:radvel}). However, since $v_i$ are outcomes of a random variable, $s_i\lam$ would still be highly  uncorrelated due to the resulting random displacements. Note that each random variable is defined at a specific pixel, over various observations, as defined in \cref{sec:random_process}.

$t_i$ on the other hand, are composed of a set of absorption lines with different strengths, but all happening at pre-defined specific wavelengths. This feature makes them highly correlated across multiple observations and easy to fit for any model -- \cref{fig:independence}, left column.

To decrease the existing correlation between realisations of the telluric signals, we increase the entropy across $t_i$ by randomising them using an emulated radial velocity, 

\begin{equation}
\begin{split}
    t_i'(\lambda) = \Vop{\tto_i(\lambda)}{v_i'}
\end{split}
\end{equation}
\noindent where each $v_i'$ is a random velocity and is drawn from 
\begin{equation}
\label{eq:randvel}
V':\mathbb{R}\rightarrow\mathbb{R},
\end{equation}
\noindent a real-valued random variable generator with an arbitrary distribution  -- which in our experiments was chosen to be uniform. The above effect can be implemented by simply contracting or expanding the wavelength axis, according to \cref{eq:radvel}. Note though that this is a purely artificial phenomenon, and although it is chosen to have the same effect as the ``real'' radial velocity of stars, it has no particular meaning for telluric lines -- a similar deterministic effect occurs when an observed spectrum is transformed into the barycentric frame.

In practice, however, we only have access to the observed signal, $x$, and not its forming components. Hence the proposed randomisation cannot be applied directly on $t$ alone, and the whole observed signal gets affected. But in the below, we show that it can still have the desired effect:

\begin{equation}
\lambda \longrightarrow \lambda_i'=\lambda(1-\frac{v_i'}{c})
\end{equation}

\begin{equation}
\label{eq:apply_velocity}
\begin{split}
    x_i'(\lambda) &= 
    \Vop{x_i\lam}{v_i'}
    \\
    &=
    \Vop{
    \Big( s_i\lam  t_i\lam \Big) 
    *h\lam
    +n_i\lam
    }{v_i'}
    \\
    &=
    \Vop{
    \Big( s_i\lam  t_i\lam \Big) 
    *h\lam
    }{v_i'}
    +\Vop{n_i\lam}{v_i'}
\end{split}
\end{equation}
\noindent which, according to the proof provided in \cref{app:conv_vop} becomes:
\begin{equation}
\label{eq:apply_velocity_2}
\begin{split}
    x_i'(\lambda) &= 
    (1-\frac{v_i'}{c})
    \Vop{
    s_i\lam  t_i\lam
    }{v_i'}
    *\Vop{h\lam}{v_i'}
    +\Vop{n_i\lam}{v_i'}
    \\
    &=
    (1-\frac{v_i'}{c})
    \Big( s_i\lamp  t_i\lamp \Big)
    *h\lamp
    +n_i\lamp\text{.}
\end{split}
\end{equation}
\noindent Now since $\left(1-\frac{v_i'}{c}\right)$ is, for each $x'_i$, a constant coefficient and is normalised out before being passed to the next step of the method, we can see that the emulated randomised velocity has found its way from $x_i$ down to $t_i$. In other words, we have achieved the required velocity randomisation in $t_i$ by modifying $x_i$. The two main components in the \textit{tweaked} signal can now be rewritten as:
\\
\begin{equation}
\begin{array}{@{}cc@{}}
\text{Telluric:} & \text{Stellar:} \\
\Vop{\tto_i\lam}{v_i'} & \Vop{\so_i\lam}{v_i+v_i'}
\end{array}
\end{equation}
\vspace{0.4cm}

\noindent So, we have achieved two components which are virtually moving independently of each other, when going over various observations. \Cref{fig:independence} illustrates this effect in practice. 

From a statistical point of view, what we achieve with the above velocity randomisation is a degree of ``whitening'' of the telluric's random process, $\{T\}$. More precisely, we decrease the mutual correlation between every pair, $(T^l,T^m)$, pushing the statistical behaviour of the telluric component toward white noise (\citealp{li1998sphering,eldar2003mmse,kessy2018optimal}). Note again that $(T^l,T^m)$, the random variables we try to decorrelate, are each defined on a single pixel of the spectrum, and their outcomes vary along with different observations. A visualisation of the results of the above whitening process is illustrated in \cref{fig:whitening}.

\begin{figure*}
  \begin{center}
    \includegraphics[width=\linewidth]{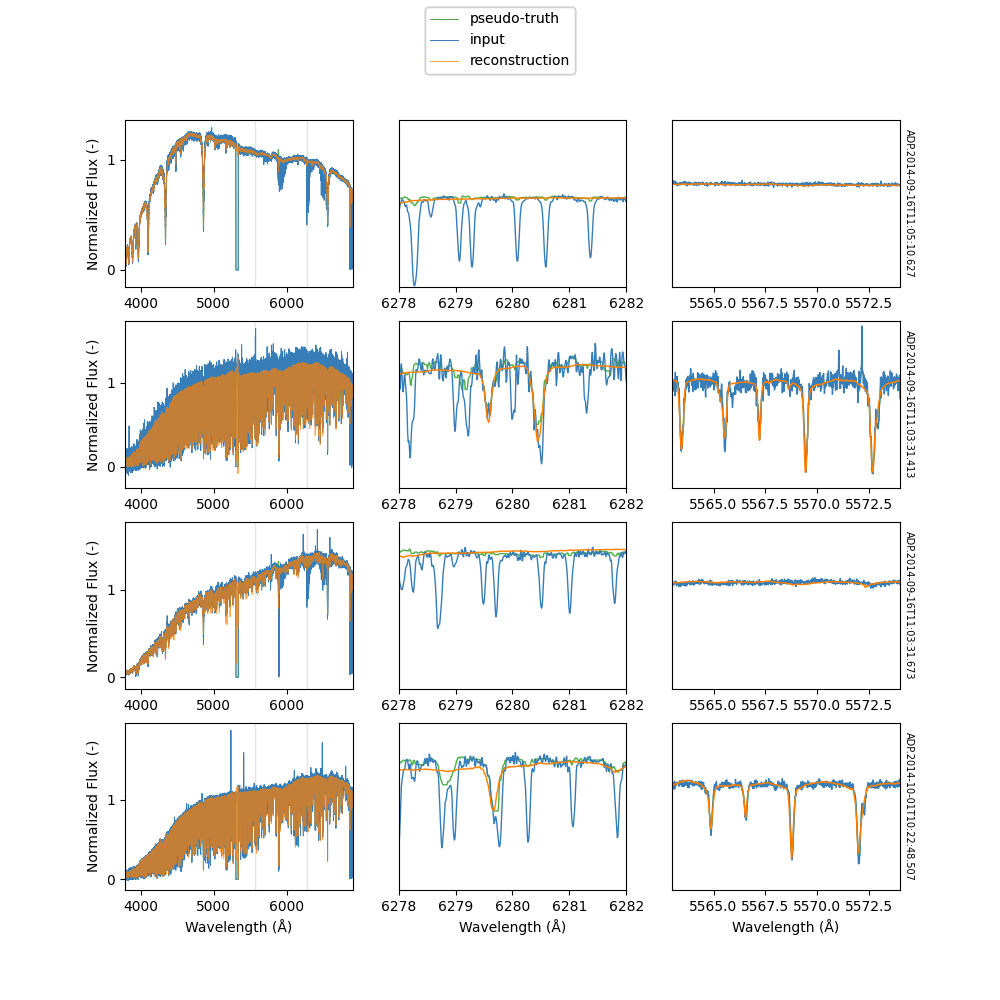}
  \end{center}
  \caption{Qualitative illustration of the results on an exemplar selection of HARPS spectra. Each row depicts one spectrum, with its HARPS ID written on the right, while columns focus on different regions of interest (indicated as grey regions on the left). Interpretation hint: reconstruction (orange) should follow the pseudo-truth (green -- see \cref{sec:quant_eval} for definition).
  The left-most column covers the whole spectrum, as is fed into the network, and illustrates the robustness of the network to different characteristics of the spectra (continuum, noise level, etc.). Major stellar lines such as $H\alpha$ can be easily spotted in the less `busy' examples.
  In the middle column we zoom in on a potentially complex region where narrow stellar lines, when existing, can collide with telluric lines of similar shapes. E.g. in the second and 4th row there are examples of such cases, where the network rejects the telluric component, while still preserving the stellar part very well -- thus rejecting the hypothesis that it might be simply rejecting narrow lines by a applying a moving average.
  In the third column, we focus on a region where  stellar lines occur in some of the spectra, but not in the others. This way we reject the hypothesis that the network might have `memorised' the locations of the lines.
  }
  \label{fig:mainresults}
\end{figure*}

\subsection{Deep Feature Space}
We borrow and use the exact architecture of the 1D convolutional autoencoder introduced by \citet{sedaghat2021machines} -- \Cref{fig:architecture1}. The convolutional layers of the \textit{encoder} transform the input spectrum down to a pre-defined number of ``latent variables'' at the bottleneck of the network. This low-dimensional representation of the input, also known as the ``code'', is the most compressed version of the input spectrum. The dimensionality of this vector is chosen based on the desired compression rate in various experiments. Note though that in the VAE version of our networks, which is the case in most of the experiments of this work, the latent nodes are implemented probabilistically, each being modelled with a pair of scalars: mean and standard deviation of a normal distribution. On the other side of the bottleneck, the decoder receives the compressed code and takes it step-by-step up to the same dimension as the original input ($2^{18} + 2^{16}$ for HARPS).
We use the same per-pixel L1 end-to-end loss function, as introduced in the original work, to achieve acceptable pixel-level accuracy.

This typical architecture has proven to be able to transform the input to a space where noise-like components of the signal are easily separable. As we show with our experiments, the novel statistical technique developed in the previous section pushes the telluric components farther from the stellar features and closer to the noise, in the learnt feature space, letting them be rejected as easily as noise.

We of course need to constrain the reconstruction abilities of the network with a high compression rate as well as a latent loss at the bottleneck, not to have too strong of a network capable of fitting the two independent components at the same time. 

Training a typical network for 100 epochs on HARPS data (number of times all the training samples are seen by the network during the training phase), using an Nvidia V100\footnote{A card with 16GB of GPU memory.}, roughly takes 60 hours. We stop the training when the validation loss plateaus.

\section{Data}
We use data from two publicly available collections: HARPS and SDSS. Each data collection calls for its own network that is trained from scratch\footnote{We loosely use the term `data collection' to refer to all the spectra collected by either a specific instrument (e.g. HARPS) or a survey (e.g. SDSS). Therefore our model that is trained on HARPS, is applicable on any new stellar spectrum that is captured by HARPS even \textit{after} the date we have collected our training set.}.
Our main experiments are run on HARPS, while SDSS is used as a difficult collection for comparison purposes.

Unlike typical ML implementations, our method, originated from archival data exploration, technically does not \textit{need to} generalise to unseen data: one can, if needed, train it on a whole set of spectra with the aim of processing the same set, which has been used during the training phase. For this particular application, since no objective for the high-level task of telluric line rejection is used, the network cannot overfit \emph{to the task}. Yet, we do want the method to generalise to unseen data. So for each data collection, we keep a small portion of samples separate from the training set, for validation purposes and to make sure that the network is not overfitting to the lower-level task of reconstruction -- i.e. memorising the spectra.
The train-validation splits are published along with our code.

\subsection{HARPS}
HARPS 
is an instrument on the 3.6 m La Silla Telescope. It is a fibre-fed high-resolution Echelle spectrograph dedicated to the discovery of exoplanets, with a spectral resolution of $R=115,000$ and covers the spectral range 378–691 nm\footnote{\url{http://archive.eso.org/wdb/wdb/adp/phase3_main/form}}. The data used has been instrument-corrected and detector-corrected, as well as sky-subtracted.

Our downloaded dataset initially consisted of 272376 total spectra, which after automatic removal of corrupted files, undefined values (NaN: Not a Number) and noise-like ones, was reduced to 267361 ``stable'' ones. This collection is mainly composed of stellar spectra. However, there are some spectra of random irregular target objects too, such as SUN, MOON, etc., which we allowed to enter our training set on purpose, to increase robustness of the learnt features. We trimmed, re-gridded and homogenised all spectra before being passed to the network in exactly the same way as is done by \cite{sedaghat2021machines} , 327680 pixels with a uniform resolution of 0.01 Å.

Note that the pipeline the HARPS spectra go through is set up such that the spectra are automatically transformed to the barycentric reference frame and re-gridded before being stored in the archive. The originally captured version, in the topocentric reference frame, is also not preserved. Therefore we had to transform them back to the topocentric frame for our experiments. Although this is touching the core concept of our presented method, it turned out to be a safe procedure: transformation to the barycentric frame exerts added randomness on the radial velocity, which is perfectly compatible with our method. In fact, our method has been inspired by observing traces of the above-mentioned fact in our initial experiments.

\subsection{SDSS}
\label{sec:sdss-data}
Our second dataset consisted of spectra from the SEGUE \citep{segueSDSS} and SEGUE-2 spectroscopic surveys \citep{segue2SDSS} that were part of the larger Sloan Digital Sky Survey (SDSS). The SEGUE surveys both used the 2.5m Sloan Foundation Telescope \citep{sloanTelescope} located at Apache Point Observatory with the two original Sloan Digital Sky Survey fiber spectrographs \citep{sloanSpectrographs}. The SDSS spectrographs have a resolution of \textit{R} $\sim 1800$ and together have 640 fibers (320 each) that plug into aluminum "plug plates" for each observation. In each plate 32 plugs are reserved for blank sky observations and 16 for spectrophotometric standard stars in the field. Each spectrograph has a red and a blue channel that collect data on separate CCDs with the blue wavelength range from approximately 3800 \r{A} to 6100 \r{A} and the red wavelengths spanning approximately 5900 \r{A} to 9200 \r{A}. Sources in the original SEGUE survey sampled Milky Way stars at a variety of distances, colours and metallicities while the SEGUE-2 targets focused on stars in the Milky Way halo.

In our experiments we used the \ntild 660,000 uncalibrated spectra from the red CCD of one of the spectrographs (labelled as 'r1' in the SDSS data archive) provided with SDSS Data Release 17 \citep{sdssDR17} and accessible via the DR17 FITS website\footnote{https://data.sdss.org/sas/dr17/}. The uncalibrated spectra consist of multiple 10-30 minute exposures of each source. Since the SEGUE surveys imaged each source multiple times to create coadded spectra this means we have multiple spectra in our dataset for the same source. For each plate we exclude the fibres that were intentionally pointed at empty patches of sky and labelled "SKY" in the data. Finally, just as in case of HARPS, we homogenise the spectra before passing them to the network. In this case, however, due to the low resolution, we end up having 2048 pixels for each spectrum, with a uniform resolution of 1.66 Å.

The uncalibrated spectra used are labelled `spFrame' in the SDSS data model\footnote{https://data.sdss.org/datamodel/} and according to the details of the SDSS spectroscopic pipeline \citep{sdssPipeline} they are flat-fielded but not flux-calibrated. The flux calibration spectra (`spFluxCalib` in the data model) include telluric absorption calculated by the spectroscopic pipeline based upon the spectrophotometric standard stars that are included in the observation set of each plate. These calibration are what we use as a "pseudo-truth" for comparisons of our network results.

\section{Experiments and Results}
\label{sec:results}
We train and evaluate multiple networks based on various combinations of the below (hyper-)parameters:
\begin{itemize}[label=\textbullet]
    \item Latent-space dimensionality
    
    \item Velocity randomisation level
    \item {Use of a variational loss}
\end{itemize}

The HARPS dataset is our primary focus and our main experiments, which include hyper-parameter sweeping and controlled tests are solely performed on this dataset. The SDSS dataset is only used for comparative testing of the method in extreme conditions -- i.e. very low resolution. We also compare our results to the tool called \texttt{wobble} \citep{bedell2019wobble} in \cref{sec:wobble_compare}.

Also, unless otherwise specified, the networks have had 128 latent dimensions, the random velocities have been drawn from a uniform distribution, roughly inspired by the tangential velocity of earth in the barycentric coordinates:
\begin{equation}  
V \sim U(-30 \, \text{km/s}, 30 \, \text{km/s}),
\end{equation}
\noindent the test set has been a fixed subset of HARPS spectra\footnote{1000 spectra of unique objects. The list is available to the public as part of the released code.}, and the region of interest (ROI) for obtaining the quantitative metrics has been $[5800, 6910]$ \AA, where there is a high density of both stellar and telluric lines.

\subsection{Primary Results}
\Cref{fig:mainresults} illustrates a few examples of how the method manages to reject the telluric lines and preserve the stellar features. The example spectra and regions are specifically chosen to rule out potential naive hypotheses for how the network rejects telluric lines. In other words, the results confirm that the network has learnt a semantic representation of the constituent components and is separating the sources in that feature space, as opposed to e.g. simply having ``missed'' narrower lines (simple averaging, low-pass filtering). In the caption of \cref{fig:mainresults} a few other such hypotheses are elaborated and rejected. The pseudo-truth used in this illustration is explained in the following subsection.

\subsection{Quantitative Evaluation}
\label{sec:quant_eval}
Various (hyper-)parameters, such as the degree of compression during dimensionality reduction, {or the degree of velocity randomization, may influence the network's ability to reject the telluric lines. There is also often a trade-off between telluric-rejection and stellar-reconstruction.

Therefore, we develop the below mutual metrics to quantify both aspects at the same time:

\begin{equation}
\label{eq:metric1}
     Q^t = \cfrac{\sum\left(M^t\left| \hat{S}-\Tilde{G} \right|\right)}
     {N^t}
\end{equation}
\begin{equation}
\label{eq:metric2}
     Q^s = \cfrac{\sum\left(M^s\left| \hat{S}-\Tilde{G} \right|\right)}
     {N^s}
\end{equation}

\noindent where $Q^t$ is a proxy for the quality of rejection of telluric lines and $Q^s$, conversely, represents the quality of stellar reconstruction. Since they are both distance metrics, lower values are desirable.
$\hat{S}$ represents the reconstructed spectrum (i.e. the direct output of the network) and $M^t$ is a binary mask; a vector of the same size as the input, having ones at all pixels containing \textit{known} telluric lines, and zeros everywhere else. $M^s$ is the dual of the tellurics mask and is simply obtained by negating it: $M^s = 1 - M^t$

\begin{figure}
    \includegraphics[width=\linewidth]{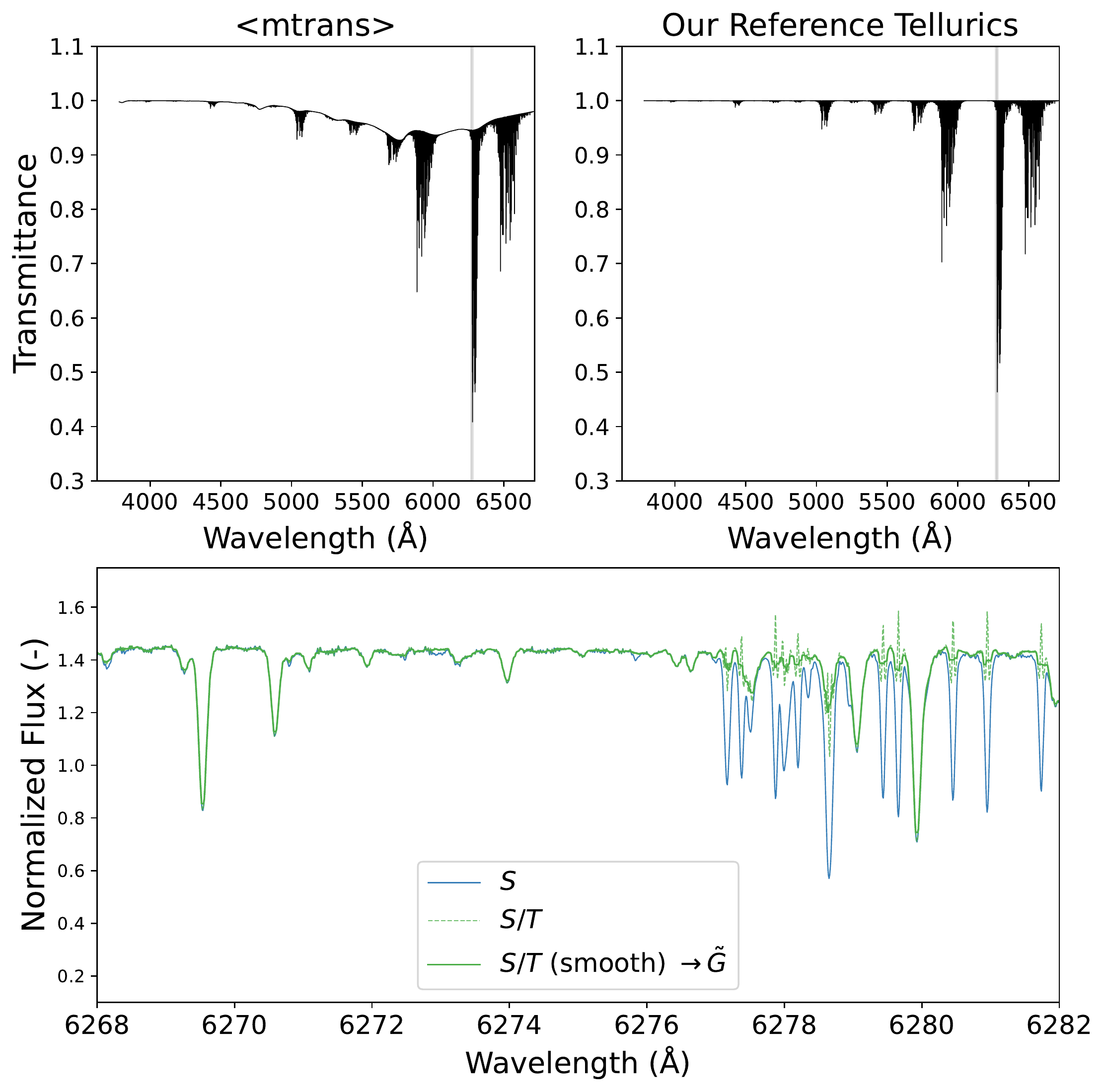}
    \caption{On the top left, the \texttt{mtrans} signal for one observation of the 51 Pegasi is illustrated (HARPS ADP ID: \texttt{ADP.2014-10-06T10:07:52.453}). The image on the top right is the same signal, after removal of the slow continuum absorption component, which forms our reference tellurics profile (T). In the bottom image, three different steps for generating the pseudo-truth spectrum are illustrated: the input spectrum (S), the result of dividing it by the tellurics profile (S/T), and finally the smoothed version. }
    \label{fig:mtrans}
\end{figure}

\paragraph*{The Pseudo-Ground Truth}, $\Tilde{G}$, is the key component of our quantitative evaluation. An ideal ground-truth for the above formulation would mean a spectrum which exactly matches the input spectrum in all aspects, but the telluric lines: same redshift, emission, absorption, CCD noise, mirror defects, but \textit{without} the telluric lines. Such an ideal spectrum would be necessary for every single object in the data collection. Obtaining such a set of signals for \textit{real} spectra is close to impossible. Note also that a mere list of wavelengths of telluric lines would not suffice for this purpose, as in the above formulation we compute pixel-by-pixel distances.
Therefore, we define the pseudo-ground truth, as a close-to-ideal truth, based on the database of lines provided by HITRAN \citep{rothman2009}. We utilise the tools provided by the \texttt{molecfit} package to access the database, taking into account the observation parameters of each spectrum, as well as the estimated density of molecules. Specifically, the \texttt{mtrans} component produced by \texttt{molecfit} encapsulates the realisation of the atmospheric transmission spectrum for the observation at hand -- see \cref{fig:mtrans} for an exemplar illustration of such a spectrum. We remove the slow, continuum absorption components \citep{rosenkranz1998water, baranov2008water}, by passing the signal through a high-pass filter\footnote{This ``continuum normalisation'' code snippet can be found in the publicly available code repository of this project.} to obtain a reference profile of telluric lines.
Now to obtain $\tilde{G}$, we first divide the input spectrum by the obtained reference profile of tellurics and then pass it through a median-smoothing kernel.
The latter is done to remove both noise and the spikes that appear at the aperture of lines, as a results of the unstable arithmetic task of division.

An example of the process leading to $\tilde{G}$ is illustrated in \cref{fig:mtrans}.

\begin{figure}
    \centering
    \includegraphics[width=\linewidth,trim=0 0 0 40,clip]{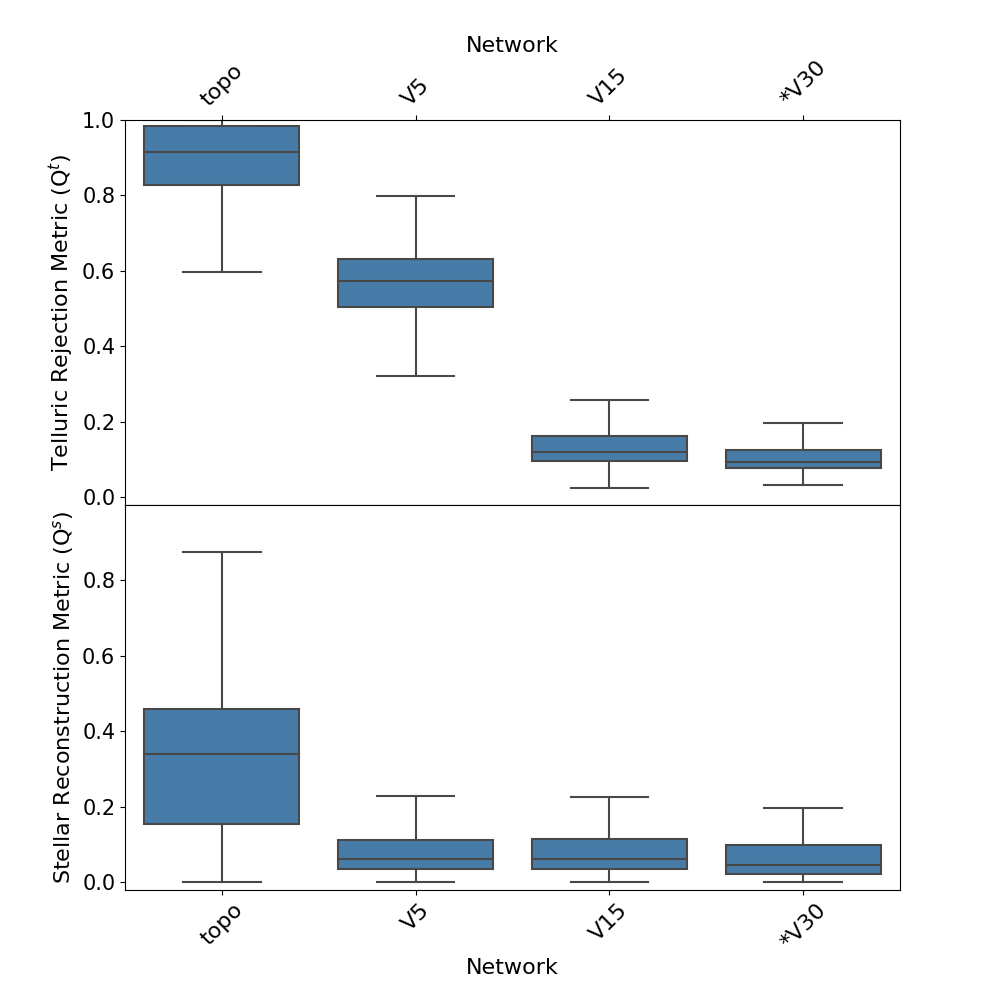}
    \caption {Illustration of the correlation between the whitening level (velocity randomisation) and tellurics rejection quality. Randomisation is set to zero in the leftmost item (`topo'), but is increased in the ones to the right of it. The numbers define the extrema of the random velocity ($\mathcal{N}$ in \cref{eq:randvel_uniform}) in km/s: e.g. V15 means $\mathcal{N}=15$ km/s and so on. The right-most item, indicated by a $*$, is our primary network, where the tellurics rejection quality has also plateaued. Note that there is not a significant difference in the stellar reconstruction quality across the randomised networks. The high variance in the \textit{stellar reconstruction} quality of `topo' may be explained by the fact that part of the learning capacity of the network is dedicated to learning to reconstruct the telluric lines.
    }
    \label{fig:compare_randvel}
\end{figure}

\paragraph*{The Normalisers}, $N^t$ \& $N^s$ in \cref{eq:metric1,eq:metric2}, are chosen such that the metrics are meaningful even when running cross-region or cross-dataset comparisons. Specifically, $N^t$ is obtained by replacing the reconstructed signal with the input and computing a \textit{unnormalised} distance metric:
\begin{equation}
     N^t = {\sum\left(M^t\left| S-\Tilde{G} \right|\right)}.
\end{equation}
\noindent The idea is that the worst outcome is when nothing has been done for removal of the telluric lines: the untouched input signal. This is of course not a very accurate assumption as there are irregular cases that can push the metric to above 1. We simply clip the value to $[0,1]$ in such cases. For $N^s$, we calculate an estimate of the continuum, $\hat{C}$, of $\tilde{G}$ to represent the worst case in which all the stellar lines have been rejected, and plug it into the (unnormalised) metric formula\footnote{This ``continuum estimation'' code snippet can be found in the publicly available code repository of this project.}:

\begin{equation}
     N^s = {\sum\left(M^s\left| \hat{C}-\Tilde{G} \right|\right)}.
\end{equation}
\noindent This way, the normalised metrics are mostly independent of the length of the considered region of interest (ROI), or the density of lines in there. Note though that, as explained before, the metrics are not designed to be robust to irregular selections of regions such as line-free areas.
In all the equations of this subsection, the subscripts $i$, indexing the observed spectra, are omitted in $S$, $\hat{S}$, $\Tilde{G}$ , $\hat{C}$, $M^s$ and $M^t$, for the sake of simplicity and the summations are calculated over all the pixels of each spectrum.

\begin{figure}
    \centering
    \includegraphics[width=\linewidth,trim=0 0 0 45,clip]{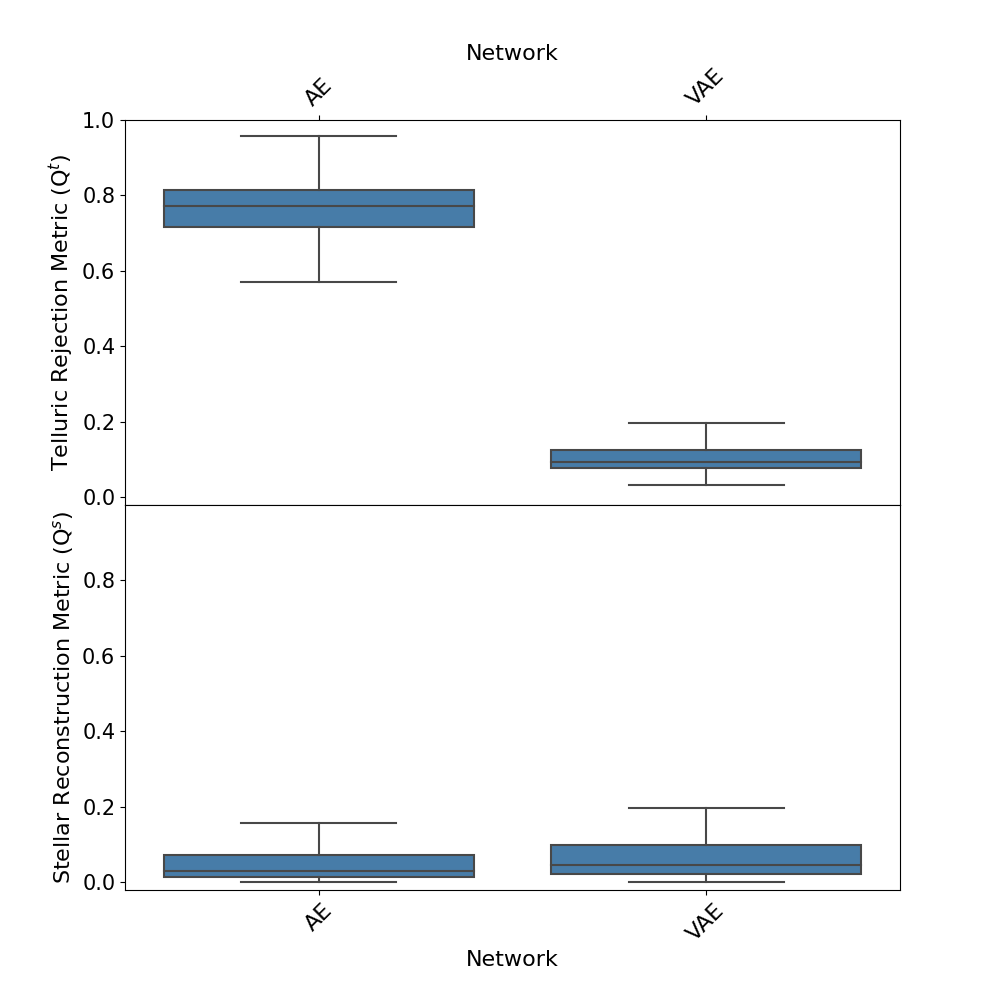}
    \caption {Quantitative comparison of a deterministic autoencoder (AE, left) with its variational couterpart (VAE, right). In the VAE, the weight of the latent loss is set to 0.1.}
    \label{fig:compare_ae}
\end{figure}

\begin{figure}

    \centering
    \includegraphics[width=\linewidth,trim=0 0 0 38,clip]{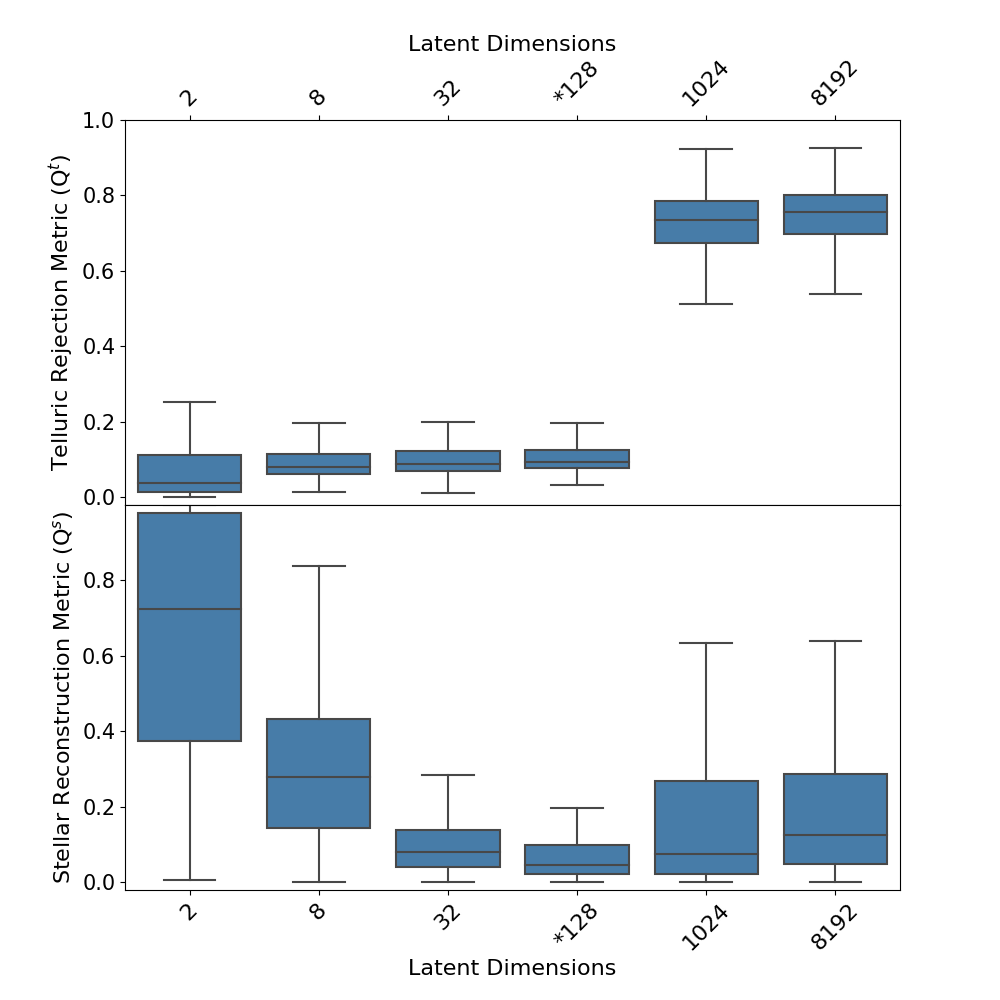}
    \caption {\label{fig:dimensionality} Quality of stellar line reconstruction vs telluric line rejection on HARPS, for different number of dimensions of the \textit{code}, or the latent representation. A lower number is better in both plots. Networks with 1024 latent dimensions and above have clearly started to learn to reconstruct telluric lines too. Our primary network ($*128$) has been chosen roughly at the optimal point, considering $Q^t$ and $Q^s$ at the same time.}
    \label{fig:enter-label}
\end{figure}

\subsection{The Effect of Velocity-Whitening}
To demonstrate the effect of the whitening step, we train and compare multiple networks in controlled conditions. We start with a network trained on spectra in the topocentric frame, representing zero velocity randomisation. Then continue to train networks with increased levels of randomisation. Concretely, we define $V'$ of \cref{eq:randvel} as a random variable with a uniform distribution:
\begin{equation}  
\label{eq:randvel_uniform}
V' \sim U(-\mathcal{N},\mathcal{N})
\end{equation}

\noindent where $\mathcal{N}$ represents the allowed maximum absolute random velocity. We increase $\mathcal{N}$ in three steps across three different networks, while other configurations are kept fixed. In \cref{fig:compare_randvel} the correlation between the randomisation level and telluric rejection is evident.

\subsection{Variational vs Deterministic Autoencoder}
A deterministic (i.e. non-variational) autoencoder, normally tends to give a better reconstruction quality at the cost of learning entangled features -- e.g. see \citealp{sedaghat2021machines}. In this work, the study of disentangled features is not the focus. But through our experiments we found that the VAE component acts as a regulariser for the network, and prevents it from learning to fit the deeper telluric lines -- i.e. memorising them. In \cref{fig:compare_ae} a deterministic autoencoder (AE) is compared to its counterpart VAE. As expected, the AE gives a slightly better reconstruction quality, while it significantly falls behind the VAE in rejection of the tellurics.

\subsection{The Effect of Compression}

It is reasonable to expect, and is supported by our experiments that, with a lower dimensionality at the information bottleneck of the network, the network's capacity for preserving many concurrent features during compression diminishes. This results in a higher rejection of \textit{details} in the reconstructed spectra, which can be naively interpreted as an improvement in telluric line rejection -- a decrease in $Q^t$. Note however that stellar reconstruction quality can get worse at the same time, and so the two metrics need to be considered at the same time.

\Cref{fig:dimensionality} compares results of various runs with different dimensionalities at the latent space. Note how the stellar reconstruction quality starts to decay by increasing the compression (lower latent dimensions). On the other hand, by increasing the number of latent nodes, the network starts to become too powerful, managing to reconstruct both components and losing the source separation capability.

\vspace{0.2cm}

\begin{figure}

    \centering
    \includegraphics[width=\linewidth,trim=0 0 0 30,clip]{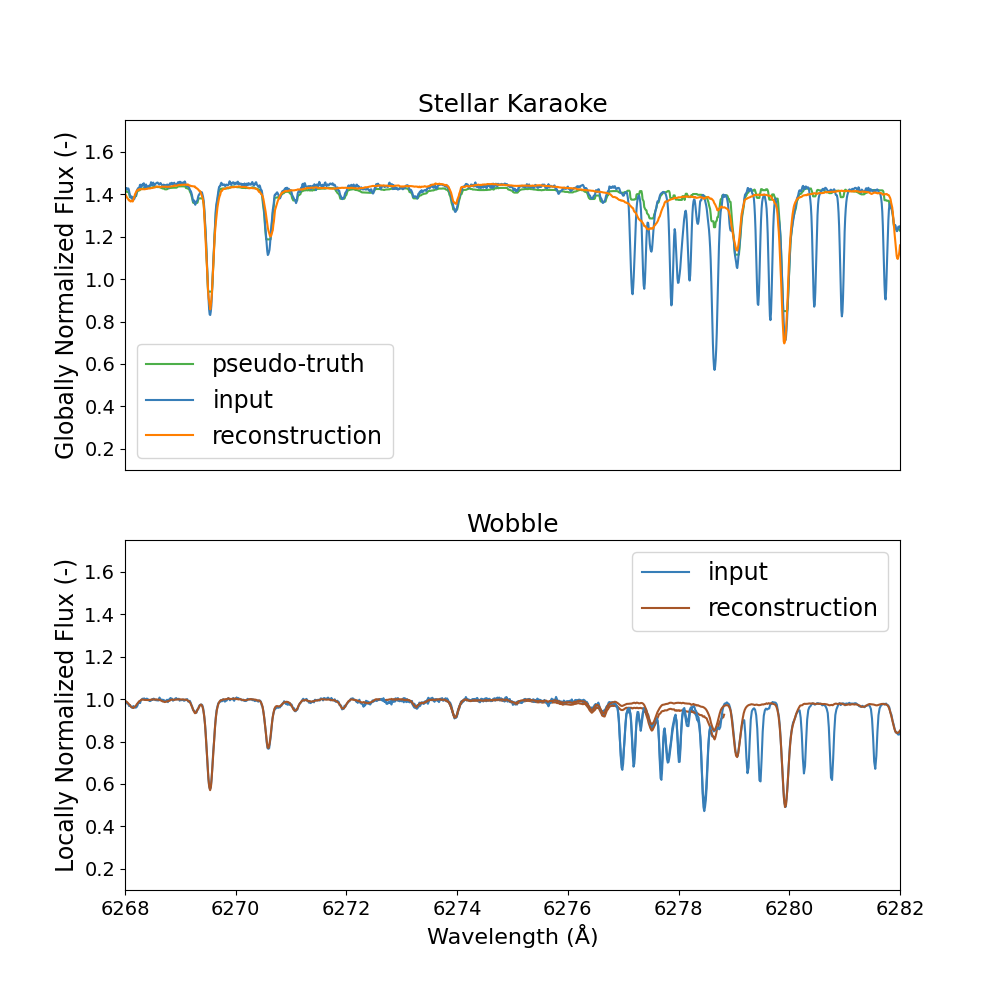}
    
    \caption {\label{tab:wobble} Spectral observations of 51 Pegasi are passed through \texttt{wobble} and Stellar Karaoke and the results are compared. In green, we have the pseudo-truth used by our method. The depicted ROI is the one used in our other visualisations, consisting of both stellar and telluric lines. For this we needed to merge two separate outputs (orders) from \texttt{wobble}: orders 62, 63.}
    \label{fig:wobble-comparison}
\end{figure}

\begin{figure}

    \centering
    \includegraphics[width=\linewidth]{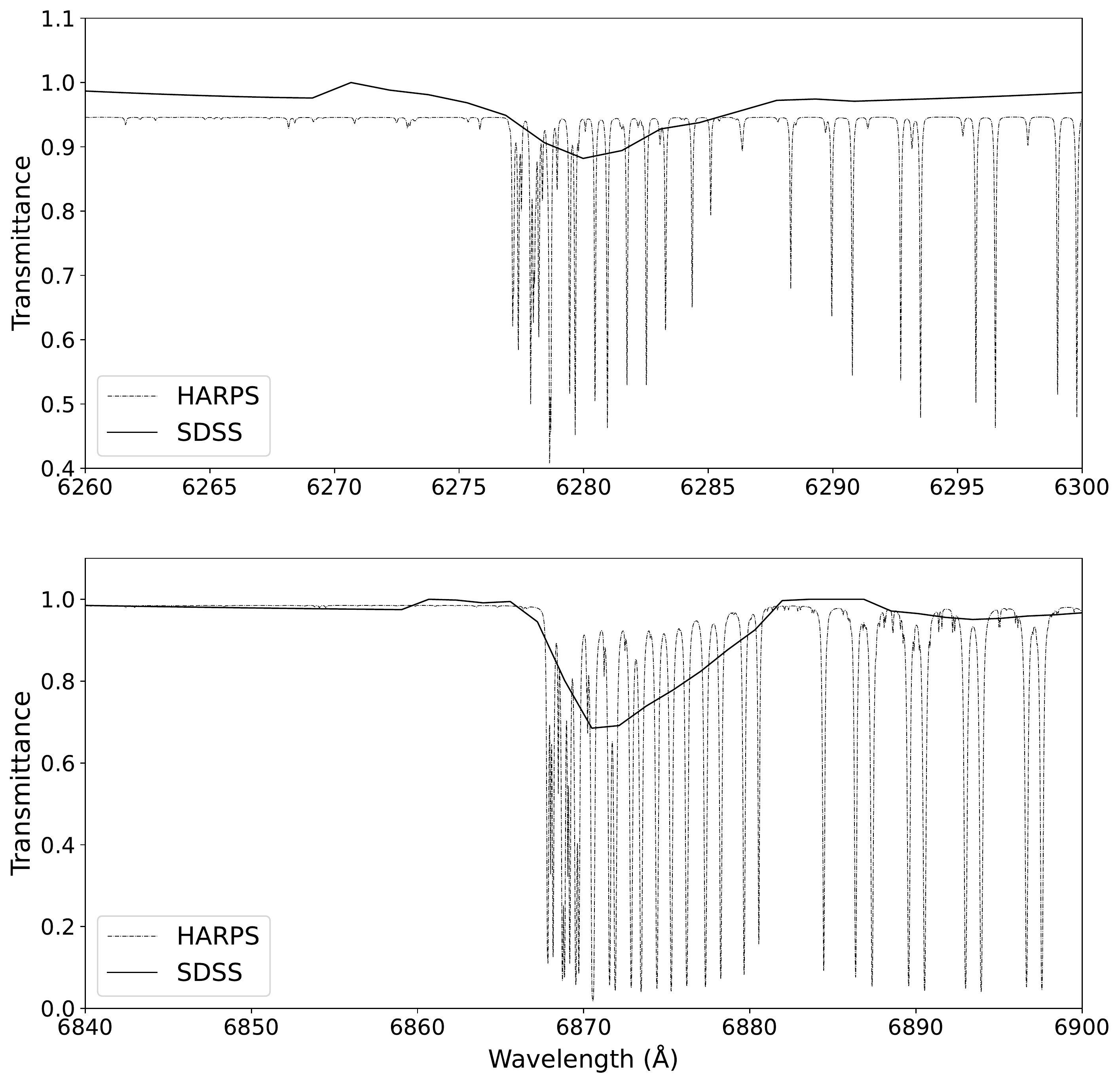}
    \caption {Telluric features are compared in a high-resolution collection (HARPS) vs a rather low-resolution one (SDSS). The two plots depict two different regions of the tellurics reference profiles, selected from the telluric-rich overlapping areas of the two collections. Most of the ``information'' in the features is lost in the low-resolution profiles. In SDSS, the tellurics profile is obtained in the same way as in HARPS, except we directly use the flux calibration provided by the survey, rather than extracting them through the \texttt{molecfit} package. See \cref{sec:sdss-data} for details.}
    \label{fig:tellu-profile-comparison}
\end{figure}

\subsection{Comparison to \texttt{wobble}}
\label{sec:wobble_compare}
The model used by \texttt{wobble} \citep{bedell2019wobble} for fitting the observed spectra has two main components: stellar vs telluric. In this sense it is one of the best candidates for comparison to Stellar Karaoke. However, it is object-specific, in the sense that it uses multiple observations of the same astrophysical object to optimise a model dedicated to that specific object. This limits the available \texttt{wobble} results to a few objects, not allowing for a statistically meaningful comparison over a large number of spectra. 
Moreover, e.g. in case of HARPS, \texttt{wobble} fits 72 different models to small wavelength spans corresponding to various echelle orders. 

These make it difficult to have a like-for-like comparison of the results. Nevertheless, in \cref{fig:wobble-comparison} we illustrate a visual comparison of both methods on 51 Pegasi\footnote{To the best of our knowledge, finding an exact correspondence between HARPS IDs and the spectra used by \texttt{wobble} is not trivial. We picked \texttt{ADP.2014-10-06T10:07:52.453} which is simply one of the observations of 51 Pegasi for this comparison. Similarly we used the first `epoch' from the pre-computed published results of \texttt{wobble}.}.
A quantitative comparison is not possible, since \texttt{wobble} follows a per-order normalisation of the flux values and performs continuum normalisation, while our method does global median-based normalisation. Qualitatively though, Stellar Karaoke and \texttt{wobble} have both rejected most of the telluric lines, as compared to the pseudo-truth -- e.g. between 6278{\AA} and 6281{\AA}. \texttt{wobble} outperforms Stellar Karaoke in following the input spectrum in shallow lines e.g. between 6271{\AA} and 6276{\AA}. In the region around 6278 \AA, the overlapping of two orders in \texttt{wobble} can be seen\footnote{See \url{https://github.com/megbedell/wobble/blob/fe32e885eba4704ff4f24345f4f99337d7d81a34/docs/quickstart.rst} for notes on current limitations in combining different orders.}.

Apart from the above mentioned methodological differences, assuming availability of multiple observations and relevant external info, \texttt{wobble} takes around 60 minutes to fit a model to an object \citep{bedell2019wobble}, whereas, Stellar Karaoke is trained only once, and then runs inference on any new spectrum in just \SI{50}{\milli\second} on a single CPU (\SI{4}{\milli\second} in case it runs on a typical Nvidia GPU).

\begin{figure*}
  \begin{center}
    \includegraphics[width=\linewidth]{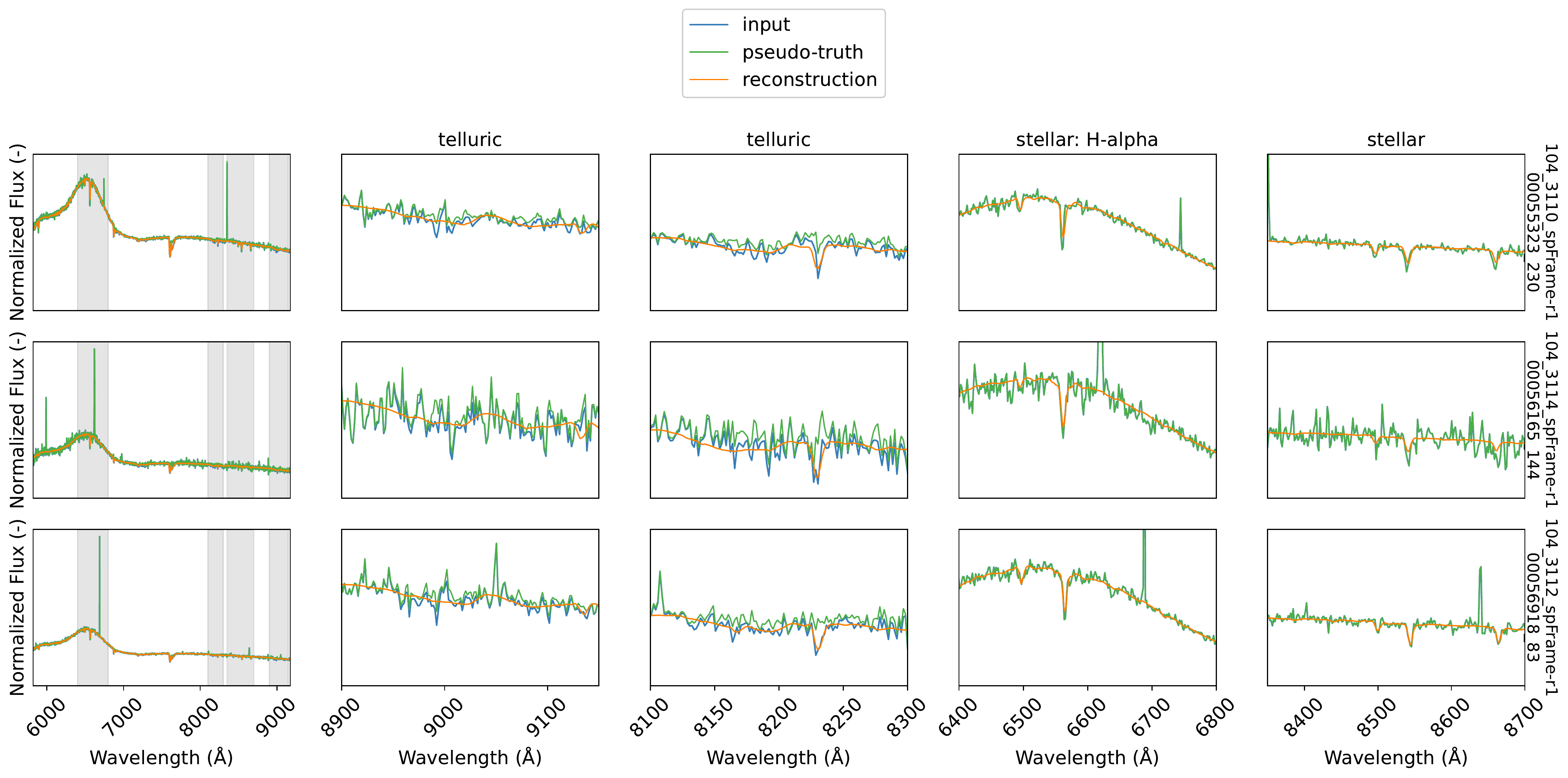}
  \end{center}

  \caption{Test results on a low-resolution data collection: SDSS. We show three random samples in the three rows. On the left, the whole spectrum is depicted, while the other columns zoom in on various regions of interest along the wavelength axis , each of which containing mainly the dominant feature specified on top of the column, 
  indicated as grey regions in the left-most column.
  Even if stellar features are preserved even in bad signal-to-noise conditions, some of the wider telluric lines are not fully rejected, due to the significantly low resolution as compared to HARPS.
  The spikes seen here are artefacts not captured by the calibration profile, hence appearing in the pseudo-truth too.
  \label{fig:sdssresults}}

\end{figure*}

\subsection{HARPS vs SDSS: High- vs Low-Resolution}
\Cref{fig:sdssresults} depicts the results of training and applying our method on SDSS spectra. The extremely low pixel resolution of this dataset (2K pixels with 1.66 {\AA} spacing, as opposed 300K pixels with 0.01 {\AA} resolution in HARPS), makes this in practice a stress test for our method, as many of the telluric lines are merged, appearing as wide artefacts -- see \cref{fig:tellu-profile-comparison} for a comparison.
  Therefore, Stellar Karaoke cannot find and learn a very meaningful representation of the telluric lines: even if the stellar lines are recovered and some of the telluric lines are being rejected, the more significant telluric artefacts are learnt and reconstructed (i.e. not rejected) by the network.
Since the quantification metrics introduced in \cref{sec:quant_eval} are mainly designed around the concept of telluric ``lines'', their values are not meaningful in this case and we determined that the qualitative comparison is conclusive.

\subsection{Comparison to \texttt{molecfit}}
\texttt{molecfit} is a widely used tool for modelling and correcting for atmospheric absorption in spectroscopic observations, particularly in case of high-resolution ones such as HARPS \citep{smette2015molecfit, kausch2015molecfit}. Since we have used a component provided by the \texttt{molecfit} package to access the HITRAN database for validation purposes, we do not provide direct comparison to \texttt{molecfit} results here. However, we list a set of methodological differences, to highlight the benefit of our method compared to \texttt{molecfit}.

\texttt{molecfit} operates by fitting synthetic transmission spectra to observed spectra, necessitating a priori knowledge of several observation-time parameters. These parameters include atmospheric pressure, temperature, humidity, airmass, observing site latitude, longitude, and altitude, time and date of observation, and instrumental setup. The accuracy of \texttt{molecfit}'s correction is, therefore, closely tied to the precision of these parameters, and any uncertainties in these values propagate into the correction process. The tool retrieves two atmospheric profiles from the Global Data Assimilation System (GDAS) website that bracket the time of the observation. These profiles are then interpolated to match the observation time and merged with a standard profile and local meteorological data to create an input profile for the radiative transfer code. \texttt{molecfit} uses a radiative transfer code to simulate atmospheric emission and transmission spectra, and it relies on a molecular spectroscopic database. In our experiments \texttt{molecfit}'s `fit-time' in average takes 3.36 cpu-minutes for each input spectrum -- this excludes any time spent on accessing databases, etc.

In contrast, an instrument-specific Stellar Karaoke network, simply receives a raw vector of scalar values as the input spectrum, and processes it in less than a second\footnote{E.g. the HARPS model presented in this paper is already trained and removes telluric lines of any new, unseen HARPS spectrum.}.

\section{Conclusions and Future Directions}
   
We presented a study that, drawing upon a Big Data-inspired view of stellar spectra, exploits the statistical independence of the radial velocity of stars with telluric lines in their observed spectra, reinforces it using a novel technique, and utilises a fully unsupervised convolutional neural network to reject the undesired part.

We see two aspects in the achievements: First, we have reported a cross-disciplinary scientific finding which exemplifies how bringing ideas from one field of science (data science), can give new insights and open doors into the other field 
of science (astronomy). 
Secondly, we have presented a method, or a prototype of it, as a promising avenue for further development when compared to traditional telluric line removal tools. 

From a practical point of view, Stellar Karaoke benefits from its ability to identify telluric lines without requiring any external information; whether it be the observation-time parameters needed in the existing methods, to manual model initialisation with lists of known telluric lines or the distribution of molecules in the earth's atmosphere in case of \texttt{molecfit} and similar methods. Training our model in practice, simply requires passing a large number of spectra through the network -- and it will do the rest. This has the potential to reduce the complexity of telluric line fitting, as well as fill in the gaps where the atmospheric and sight measurements may be missing or inaccurate.

Stellar Karaoke is extremely fast. Training one network suffices for an entire data collection, as well as any new spectrum to be added to it in the future.
Once trained, it processes each spectrum in a fraction of a second, a noteworthy benefit considering the current processing times are several minutes and tens of minutes for \texttt{molecfit} and \texttt{wobble} respectively, while today's data collections encompass hundreds of thousands of spectra.
To highlight the importance of this, we ran Stellar Karaoke on all of the approximated 260K publicly available HARPS data and made the results available to the public; to the best of our knowledge this is an unprecedented step, likely due to the previously required time and computational resources.

Nevertheless, the current version is still a demonstration of a novel research product, showcasing the strengths of a fully unsupervised approach. For Stellar Karaoke to evolve into a ready-to-use \emph{tool} in every application, additional refinements are needed. Notably, due to the unsupervised nature of the research, the network has now decided to reject the telluric component and reconstruct the stellar one. This is of course a result of the statistical nature of the two phenomena (more significance in stellar lines, etc.). But to make it a more robust tool, less sensitive to the hyperparameters involved, one can introduce a `hint' to the training process to direct the network's attention towards the desired component -- e.g. mutual information between the decomposed components.

Another avenue for exploration is the case of outliers. Like any other method, there are cases for which the method underperforms -- either completely, or to a noticeable extent. A slight degree of supervision, such as the manual injection of rare samples to the training set, can come to help in such a case.

\section*{Acknowledgements}

Some of the experiments demonstrated in this work have been run on compute servers provided by ESO, during NS's collaboration with the ESCAPE project between 2019 and 2021.
Also, authors 1-4 acknowledge support from the DiRAC Institute in the Department of Astronomy at the University of Washington. The DiRAC Institute is supported through generous gifts from the Charles and Lisa Simonyi Fund for Arts and Sciences, and the Washington Research Foundation.

\section*{Data Availability}
We have released the code for velocity whitening and the convolutional neural network, as well as the list of IDs of the spectra from both data collections, on \url{https://github.com/NimSed/stellar-karaoke}. Sample implementations and outputs, as well as potential future versions of the method will be publicly available on the project web-page: \url{https://www.rawdataspeaks.com/projects/stellar-karaoke}.

\bibliographystyle{mnras}
\bibliography{ref}

\appendix
\section{Wavelength Transformation and Convolution}
Here we show how scaling of the wavelength axis is distributed into the various components of an observed spectrum -- related to \cref{eq:apply_velocity}, \cref{eq:apply_velocity_2}.
\label{app:conv_vop}
\begin{equation}
\begin{split}
    x\lam &= z\lam * h\lam
    \\
    &= \int_w z(w) h(\lambda-w) dw
\end{split}
\end{equation}

\begin{equation}
\begin{split}
    x\lamp &= x\alam
    \\
    &= \int_w z(w) h(a\lambda-w) dw
\end{split}
\end{equation}

\begin{equation}
    \text{Let\ }
    w = au,
\end{equation}

\begin{equation}
\begin{split}
    x\lamp
    &= a \int_u z(au) h(a\lambda-au) du
\end{split}
\end{equation}

\begin{equation}
    \text{Let\ }
    z(au)=z'(u),\ h(au)=h'(u)
\end{equation}

\begin{equation}
\begin{split}
    \Longrightarrow 
    x\lamp
    &= a \int_u z'(u) h'(\lambda-u) du
    \\
    &= a\ z'\lam*h'\lam
    \\
    &= a\ z\alam*h\alam
    \\
    &= a\ z\lamp*h\lamp
\end{split}
\end{equation}

The same could be shown in a rather shorter way using the Fourier transform. However, to avoid confusion between the terms \textit{frequency} and \textit{wavelength} used in two different domains, here we use the direct expansion of the convolution operator.

\section{A fully-connected architecture}
\label{app:fullyconnected}
Below the details of the trainable parameters of an exemplar fully-connected autoencoder are depicted -- a hypothetical network related to \cref{sec:intro-auto-encoders}, not the network used for Stellar Karaoke. To bring down the dimensionality of a HARPS spectrum, even in only three steps, the number of trainable parameters quickly goes beyond 6B; a dramatic difference compared to the convolutional counterparts.
\begin{verbatim}
==============================================
Layer (type:depth-idx) Output Shape Param #
==============================================
AE1D_FC               --           --
|--Sequential: 1-1     [1, 1, 128]  --
|  |--Linear: 2-1       [1, 1, 10000] 3,276,810,000
|  |--ReLU: 2-2         [1, 1, 10000] --
|  |--Linear: 2-3       [1, 1, 1000]  10,001,000
|  |--ReLU: 2-4         [1, 1, 1000]  --
|  |--Linear: 2-5       [1, 1, 128]   128,128
|  |--ReLU: 2-6         [1, 1, 128]   --
|--Sequential: 1-2     [1, 1, 327680]--
|  |--Linear: 2-7       [1, 1, 1000]  129,000
|  |--ReLU: 2-8         [1, 1, 1000]  --
|  |--Linear: 2-9       [1, 1, 10000] 10,010,000
|  |--ReLU: 2-10        [1, 1, 10000] --
|  |--Linear: 2-11      [1, 1, 327680]3,277,127,680
|  |--ReLU: 2-12        [1, 1, 327680]--
==============================================
Total params: 6,574,205,808
Trainable params: 6,574,205,808
\end{verbatim}

\bsp	
\label{lastpage}
\end{document}